\documentstyle[12pt,aps]{revtex}

\begin{document}
\begin{center}
{\large\bf Model for two generations of fermions}\\
\vspace*{3mm}
V.V.~Kiselev\\
Institute for High Energy Physics,\\
Protvino, Moscow Region, 142284, Russia\\
E-mail: kiselev@mx.ihep.su, kiselev@th1.ihep.su, 
Fax: +7-(095)-2302337, Phone: +7-(0967)-713780
\end{center}

\begin{abstract}
In the model with the spontaneous breaking of chiral gauge symmetry,
the vacuum structure for the pair of Higgs fields can provide the introduction
of two generations of fermions. The mixing matrix of charged currents is
determined.
\end{abstract}

\vspace*{4mm}

\section{Introduction}

Two generations of fermions, say, electron and muon, have identical properties
with respect to a gauge interaction and differ by the mass values yet.
In the Standard Model \cite{ws} 
the masses are caused by the spontaneous breaking
of gauge symmetry \cite{hig}, 
so that the difference between the masses corresponds
to the variation of Yukawa constants for the interaction between the
fermions and Higgs fields. The values of those constants and the origin
for the replication of generations are the problems, which are studied beyond
the Standard Model.

At present, the origin of generations is usually described \cite{pecc}
in the framework of
\begin{itemize}
\item[1.]
composite models with a discrete symmetry of preons \cite{composit}, which
results in a general form of mass-matrices for the effective fermions coupled
to the complex bosonic fields,
\item[2.]
random lattice dynamics of gauge interactions \cite{random},
\item[3.]
geometric nature of generations: superstring models \cite{GSW} in a
space of extended dimensions compactified at low energies \cite{super}.
\end{itemize}

Thus, the appearance of fermion replications is caused
by the dynamics of some extended set of fundamental fields generally including
effective Higgs fields, which have the same number of generations as the
fermions.

In this work we describe the model of fermions possessing a chiral gauge 
interaction and coupled to two Higgs fields. The latter ones have a self-action
potential with a particular properties of minimal energy corresponding 
to the special symmetry of the vacuum. So, there are only two configurations of
minimum, which are not related by the gauge transformation and give two values
of fermion masses, correspondingly. The full vacuum state including these two 
configurations is symmetric over the action of discrete cyclic group. There is
no physical principle to restrict the theory by the only configuration. Thus,
the symmetric vacuum provides the {\sf introduction} of two generations of
fermions interacting with the gauge field.

The model of fermion interaction and the Higgs potential are 
constructed in Section II. The introduction of generations is described.
The model with the charged currents is considered in Section III, where
the mixing matrix is determined. The results are summarized in Conclusion.

\section{Model}

Consider the chiral interaction of dirac spinor with the abelian
gauge field ${\cal A}_\mu$ and Higgs field $h$
\begin{equation}
L=\bar \psi_R p_\mu \gamma^\mu \psi_R +
\bar \psi_L(p_\mu - e{\cal A}_\mu)\gamma^\mu \psi_L +g\bar\psi_R \psi_L h
+g\bar\psi_L \psi_R h^* +L_{GF}+L_0(h,{\cal A}),
\label{f1}
\end{equation}
where $L_{GF}$ is the Lagrangian of free gauge field (with a possible
introduction of term fixing the gauge and the corresponding Lagrangian
for the Faddeev--Popov ghosts), $L_0$ is the Lagrangian of
Higgs field with a self-action and the gauge interaction,
$g$ is the Yukawa constant. At the spontaneous breaking of symmetry,
the vacuum expectation of Higgs field is not equal to zero
\begin{equation}
\langle 0| h |0\rangle = \eta e^{i\phi}.
\label{f2}
\end{equation}
In the unitary gauge $\phi=0$, the fermion becomes massive\footnote{
The appearance of mass for the gauge field is not an object under 
consideration here.}, so $m=g\eta$
\begin{equation}
L_m = g\eta (\bar \psi_R\psi_L + \bar \psi_L \psi_R).
\label{f3}
\end{equation}
In the arbitrary gauge, it is convenient to use the method of auxiliary
fields, so that, for example,
\begin{equation}
f_L =\frac{1}{\sqrt{2}} \left( {\psi_L}\atop {e^{i\phi}\psi_L}\right),
\;\;\;\;\;
f_R =\frac{1}{\sqrt{2}} \left( {e^{-i\phi}\psi_R}\atop {\psi_R}\right),
\label{f4}
\end{equation}
and
\begin{equation}
L_m = \bar f_R M_{RL} f_L + \bar f_L M_{LR} f_R,
\label{f5}
\end{equation}
where
\begin{equation}
M_{RL} = M_{LR} = g\eta \left( \begin{array}{cc} 
1 & 0 \\ 0 & 1 \end{array}
\right).
\label{f6}
\end{equation}
By the procedure, the kinetic term and the gauge interaction of
$f$ with the $\cal A$ field are not changed. Thus,
one can see in (\ref{f6}), that we have the field with the mass $m=g\eta$,
as one must expect from the unitary gauge (\ref{f3}).

It is convenient to introduce the following fields
\begin{equation}
\tilde f_L =\frac{1}{\sqrt{2}} \left( {\psi_L}\atop {i\psi_L}\right),
\;\;\;\;\;
\tilde f_R =\frac{1}{\sqrt{2}} \left( {\psi_R}\atop {-i\psi_R}\right),
\label{f7}
\end{equation}
so that one has 
\begin{equation}
\tilde M_{RL} = \tilde M_{LR} = 
g\eta \left( \begin{array}{cc} \cos\phi & \sin\phi \\ \sin\phi & -\cos\phi 
\end{array} \right),
\label{f8}
\end{equation}
with the eigenvalues $|\lambda|=m=g\eta$.

Now consider the analogous interaction of fermions with two higgses
$h_1$ and $h_2$. For the latter ones at the spontaneous breaking of
gauge symmetry, the vacuum expectation values are equal to
\begin{equation}
\langle 0| h_1 |0\rangle = \eta_1 e^{i\alpha}, \;\;\;\;
\langle 0| h_2 |0\rangle = \eta_2 e^{i\alpha+i\phi},
\label{f9}
\end{equation}
where $\alpha$ is an arbitrary gauge parameter,
$\phi$ is the difference between the phases of {\sc vev},
so that this difference can be observable. Further, the introduction
of fields (\ref{f7}) results in the appearance of fermion mass
$m$ depending on $\phi$
\begin{equation}
\frac{1}{g^2} m^2 = (\eta_1+\eta_2\cos\phi)^2+\eta_2^2\sin^2\phi =
(\eta_1+\eta_2)^2-4\eta_1\eta_2\sin^2\frac{\phi}{2}.
\label{f10}
\end{equation}

The following situation is of a special interest: the phase difference $\phi$ 
is constrained due to the form of higgs self-action, so that the vacuum
minimum of the potential takes a place at some discrete values of $\phi$.

In the model under consideration, the higgs potential has the form
\begin{equation}
V=\frac{\lambda_{11}}{4}\big(h_1h_1^*\big)^2-\frac{\mu^2}{2}h_1h_1^*
+\frac{\lambda_{22}}{4}\big(h_2h_2^*\big)^2
-\frac{\lambda_{12}}{8}\big[h_1h_2^*+h_2h_1^*\big]^2,
\label{v}
\end{equation}
where all of $\lambda_{ij}$ and $\mu^2$ are greater than zero, so that the
constraint
of the potential stability ($V>0$ at infinity) gives 
$\lambda_{11} \lambda_{22}>\lambda^2_{12}$.

Then the constraints on a stable minimum of static energy,
$$
\frac{\partial V}{\partial |h_1|} = 0, \;\;\;
\frac{\partial V}{\partial |h_2|} = 0,
$$
result in the vacuum expectation values equal to
\begin{eqnarray}
\eta_1^2 & = & \frac{\mu^2}{\lambda_{11}-\frac{\lambda_{12}^2}
{\lambda_{22}}\cos^4 \phi}, \\
\eta_2^2 & = & \frac{\lambda_{12}}{\lambda_{22}}\eta_1^2\cos^2 \phi.
\end{eqnarray}
Then the minimum energy is equal to
\begin{equation}
V_{min} = -\frac{1}{4}\; \frac{\mu^4}{\lambda_{11}-
\frac{\lambda_{12}^2}{\lambda_{22}}\cos^4\phi}.
\end{equation}
Therefore, in the vacuum state one has $\cos^2 \phi=1$.
Thus, in the specified model of higgs self-action the phase difference
between the vacuum expectations runs over the discrete values
\begin{equation}
\phi= \pi n,\;\;\; n\in {\rm Z.}
\label{f12}
\end{equation}
Then introducing fields (\ref{f4}), one gets
\begin{equation}
M_{RL} = M^{\dag}_{LR} = g\left( \begin{array}{cc} 
\eta_2 & \eta_1 e^{-2i\phi} \\ \eta_1 & \eta_2 \end{array}
\right),
\label{f11}
\end{equation}
and at fixed $\phi$ (\ref{f12})
the real matrix $M_{RL}=M_{LR}$ has the eigenvalues
corresponding to the fermion masses
\begin{equation}
m_{1,2} = g |\eta_1\pm \eta_2|.
\label{f13}
\end{equation}
Further, introducing
\begin{equation}
U= \frac{1}{\sqrt{2}} \left( \begin{array}{rr} 
1 & 1 \\ -1 & 1 \end{array} \right),
\label{u}
\end{equation} 
one gets the diagonal form of the mass-matrix
$$
M^U_{LR} = M^{U}_{RL} = U \cdot M_{LR}\cdot U^{\dag} = g \left(
\begin{array}{cc} 
\eta_2+\eta_1 & 0 \\ 0 & \eta_2-\eta_1 \end{array} \right)
$$
with the eigen-vectors $f^{U}_{L,R} = U f_{L,R}$.

Let us emphasize that, first, the introduced auxiliary fields have the 
particular form
$$
f^U_L= \left( \psi^{(1)}_L \atop 0 \right),\;\;\;\;\;
f^U_R= \left( \psi^{(1)}_R \atop 0 \right),
$$
at $\phi=\phi_{(1)}=0$, so that acting on the minimum-energy configuration
$|0_{(1)}\rangle$ the field $\psi^{(1)}$ results in the state with the mass
$m^{(1)}= g (\eta_1+\eta_2)$. Second, one obtains
$$
f^U_L= \left( 0 \atop -\psi^{(2)}_L \right),\;\;\;\;\;
f^U_R= \left( 0 \atop \psi^{(2)}_R \right),
$$
at $\phi=\phi_{(2)}= \pi$, so that the $\psi^{(2)}$ field has 
the mass $m^{(2)}= g |\eta_1-\eta_2|$ over the configuration $|0_{(2)}\rangle$.
Therefore, if one constructs the model vacuum as
\begin{equation}
|{\rm vac}\rangle = |0_{(1)}\rangle \otimes |0_{(2)}\rangle =
\left( |0_{(1)}\rangle \atop |0_{(2)}\rangle \right),
\end{equation}
which is symmetric over the action by the discrete second-order cyclic group,
changing the complex phase differences between the {\sc vev} of higgses,
then one can use two kinds of auxiliary fields at $\phi=0$ and $\phi=\pi$,
having the mentioned particular form and the common mass-matrix independent of
the chosen values of $\phi$, {\sf to introduce} the common {\sf physical} field
with the components corresponding to the kind of generation. So, the extended
Lagrangian contains the physical field $f$ over the vacuum $|{\rm vac}\rangle$
with two generations.

The structure of the Higgs fields vacuum
can not change the number of fermionic degrees of freedom, so, it can not be
the origin of generations, since it can not produce these fermion fields.
Two generations are introduced by construction.
However, the considered model of Higgs vacuum certainly provides
the introduction of two generations: the Higgs vacuum is arranged
for the introduction of two generations of fermions. So, the number of
fermionic degrees of freedom is not changed in an original Lagrangian,
since it is {\sf introduced} in the self-consistent way.

It is correct that in a gauge theory the only possible vacuum
configuration can be chosen because the physical meaning of gauge invariance is
the equivalence of those configurations: the physics in each configuration
is the same, and the only configuration must be chosen, the others are produced
by the gauge transformations. So, the physical principle for the
restriction by the only configuration is the gauge invariance. However, that
is not the case of the paper model under consideration. The constructed
configurations of minimal energy are not gauge equivalent, since they differs
by the phase difference which is the observable physical quantity. Therefore,
the full vacuum of the theory must contain all of configurations because there
is no physical principle ("a rule of superchoice") to cancel some configuration
and to restrict the state. So, the paper describes the theory, where all
of configurations are symmetrically presented in the vacuum.

Thus, the pair of fermion species, possessing the identical properties
over the gauge interaction and having the different masses, are introduced due
to two configurations of symmetric vacuum for the
pair of Higgs fields, so that in this model the effect of two generations
of fermions is provided by the structure of vacuum for the Higgs fields.

From the practical point of view, equation (\ref{f13}) shows
that the large splitting of fermion masses for two generations
($e$, $\mu$) can be caused by no difference between the
values of Yukawa constants for the interaction between the
fermions and higgs. The mass difference can be the result of
small splitting between the vacuum expectation values\footnote{
Supposing $m_1/m_2 = m_e/m_\mu$, one finds that $\Delta\eta/\eta \approx 1/103
\sim \alpha_{em}$, so that, probably, the splitting is of the order of
radiative corrections.}
(at $\eta=\eta_1>\eta_2>0$,
$\Delta\eta=\eta_1-\eta_2\ll \eta$)
\begin{equation}
\frac{m_1}{m_2} = \frac{\eta_1-\eta_2}{\eta_1+\eta_2} \approx
\frac{\Delta\eta}{2\eta}\ll 1.
\label{f14}
\end{equation}

Thus, we have shown that at the spontaneous breaking of symmetry
in the interaction of fermion field with the pair of
higgses, the structure of higgs vacuum can provide the introduction of
two generations of fermions.

\section{Mixing of charged currents}

The introduction of auxiliary fields (\ref{f4}) with the symmetric mass-matrix
at $\phi=0, \pi$ contains the uncertainty related with an additional
term 
$$
\Delta M = \left( \begin{array}{rc} -a & 0\\ 0 & a \end{array} \right),
$$
which does not contribute to the determination of the mass values,
since it is canceled in the Lagrangian expressed through the initial
single-generation fields. This uncertainty corresponds to the rotation of the
{\sf introduced} physical fields in the Lagrangian with two generations.
So, considering the general form of extended Lagrangian with different Yukawa
constants $g_1$ and $g_2$ for the Higgs fields $h_1$ and $h_2$, one gets the 
following expression for the mass-matrix\footnote{The number of relative
phases in the set of couplings $\{g_1,g_2,\ldots,g_n\}$ and fields $\{h_1,h_2,
\ldots,h_n\}$ is equal to $n_\delta=(n-1)(n-2)/2$, so that for $n=2$ 
we have the situation, when $n_\delta=0$, and $\{g_1,g_2\}$ are real and
positive.}
\begin{equation}
M = \left( \begin{array}{cc} v_2-v_1 \sin 2\theta & v_1 \cos 2\theta \\
v_1 \cos 2 \theta & v_2+v_1 \sin 2 \theta \end{array} \right),
\end{equation}
where $v_i=g_i\eta_i$, so that the eigenvalues of the matrix are the same 
$m_{1,2}=|v_1\pm v_2|$, which are independent of the $\theta$ value. However,
there is a particular state, when $\theta$ is determined by the
physical quantities of model. So, we find\footnote{We consider the case of
$v_2/v_1\le 1$. A description of the inverse condition is quite evident after 
a suitable transformation of $\psi_R$.}
\begin{equation}
M = \left( \begin{array}{cc} v-a & v\\ v & v+a \end{array} \right),
\label{mv}
\end{equation}
where $v=v_2$, $a=v \tan 2 \theta$, 
$$
\cos 2\theta = v_2/v_1= \frac{g_2}{g_1} \sqrt{\lambda_{12}/\lambda_{22}}.
$$ 
Note, that matrix (\ref{mv}) has the symmetric texture: 
$$
M(1,2)=M(2,1)=\frac{1}{2}[M(1,1)+M(2,2)].
$$
The $\theta$ value is related with the masses of fermions
\begin{equation}
\tan 2\theta = 2 \frac{\sqrt{m_1 m_2}}{m_1-m_2}.
\end{equation}
The matrix gets the "see-saw" form in the "heavy" basis \cite{dem}
\begin{equation}
M^{U} = U\cdot M\cdot U^{\dag} = \left( \begin{array}{cc} 2v & a\\ a & 0
\end{array}
\right),
\label{mv2}
\end{equation}
where $U$ is defined in eq.(\ref{u}). At $v_1=v_2$ we have $a=0$, and 
the mass-matrix (\ref{mv}) is a "democratic" one \cite{dem}.

Next, the model matrix  (\ref{mv2})  takes the diagonal form after
the action by the rotation to the angle $\theta$.

Let us consider now the model with two kinds of the chiral fields 
$\psi^{(u)}$ and $\psi^{(d)}$ possessing different charges, so that
the charge current interaction between the latter ones has the form
$$
L_{cc} = e \bar \psi^{(u)}_L W_\mu\gamma^\mu \psi^{(d)}_L + \mbox{h.c.},
$$
where the initial mass-matrices have the form (\ref{mv}) with
$$
\cos 2\theta^{(u)}
=\frac{g_2^{(u)}}{g_1^{(u)}}\sqrt{\frac{\lambda_{12}}{\lambda_{22}}},\;\;\;
\cos 2\theta^{(d)}
=\frac{g_2^{(d)}}{g_1^{(d)}}\sqrt{\frac{\lambda_{12}}{\lambda_{22}}}.
$$
Then the Cabibbo mixing matrix has the form
$$
V = \left( \begin{array}{rc} \cos \theta_c & \sin \theta_c \\
-\sin \theta_c & \cos \theta_c \end{array} \right),
$$ 
where $\theta_c=\theta^{(u)}-\theta^{(d)}$. At $\frac{m_2^{(u)}}{m_1^{(u)}}\ll
\frac{m_2^{(d)}}{m_1^{(d)}}\ll 1$ one gets
the Cabibbo angle
$$
\sin\theta_c\approx \; \sqrt{\frac{m_2^{(d)}}{m_1^{(d)}}}.
$$
Thus, the offered model provides the introduction
of two generations due to the structure of vacuum for
two Higgs fields. It describes also the mixing matrix of charged 
currents through the physical quantities, that allows one to relate 
the mixing angle with the masses of fermions.

\section{Conclusion}

In this paper we have constructed the potential of two Higgs fields,
which have the discrete symmetry of the vacuum state. Two observable
differences between the phases of the vacuum expectation values
exist only, which results in two possible values of mass for the fermion fields
coupled to the higgses. These two configurations of symmetric vacuum provide
the introduction of two generations of fermions.

In the presence of charged currents the mixing angle can be related
with the mass ratios determined by the coupling constants of the
model.

Certainly, the construction of a realistic model for three generations of
leptons and quarks is of a special interest beyond the scope of this paper.
Note, that at $\eta_1=\eta_2$ the mass matrix becomes symmetric under
the permutations of its elements. This texture is usually referred to the
"democracy" pointing out the equal contribution of different flavors
\cite{dem}. The form provides the observed regularity of masses for
the lepton and quark generations, as the single generation is heavy only, when
two others can be considered to be massless in the leading
approximation. An introduction of corrections breaking down the permutation
symmetry allows one to get some relations \cite{dem}
for both the fermion masses
and the elements of Cabibbo--Kobayashi--Maskawa matrix.

\end{document}